\documentclass[epj]{svjour}
\usepackage{latexsym}

\newcommand{\fant}[1]{\phantom{#1}}
\newcommand{\be}{\begin{equation}}
\newcommand{\ee}{\end{equation}}
\newcommand{\wdg}{\wedge}
\newcommand{\ot}{\otimes}

\begin{document}

\title{Linearized gravity in terms of differential forms}
\author{
Ahmet Baykal\inst{1}\thanks{\emph{abaykal@ohu.edu.tr}}
\and 
Tekin Dereli\inst{2}\thanks{\emph{tdereli@ku.edu.tr}}
}
\institute{
Department of Physics, Faculty of Arts and Sciences, \"Omer Halisdemir University, Merkez Kamp\"us, Bor yolu \"uzeri, 51240,  Ni\u gde, TURKEY
\and 
Department of Physics, College of Sciences, Ko\c{c} University, Rumeli Feneri Yolu, 34450 Sar\i yer, \.Istanbul, TURKEY
          }

\date{Received: 
/ Revised version:\today }

\abstract{
A technique to linearize gravitational  field equations is developed in which the perturbation metric coefficients are treated as second rank, symmetric, 1-form fields belonging to the Minkowski background spacetime by using the exterior algebra of differential forms. 
\PACS{
      {04.20.Cv}{ Fundamental problems and general formalism}  \\
			\fant{PACS...}{04.20.Fy}{ Canonical formalism, Lagrangians, and variational principles}\\
      \fant{PACS...}{04.50.Kd}{ Modified theories of gravity}
     } 
      } 
\maketitle

\section{Introduction}

Immediately after the formulation of the field equations for the theory of general relativity a century ago, Einstein studied the  field equations bearing his
name in the linear approximation. In particular, he derived the amount of deflection of the starlight by a spherically symmetric mass by using the linear approximation and he also introduced gravitational waves and  derived the quadrupole formula for the energy emitted by a material source in the form of gravitational waves.

The gravitational waves generated by a merger of two massive black holes has been observed for the first time  by gravitational wave  detectors of the  LIGO collaboration \cite{ligo} almost exactly  a century later. This landmark observation will certainly have an impact on the theoretical front in  general relativity. The gravitational waves topic  with all its aspects should be expected to gain  more popularity in  general relativity courses. 

The exterior algebra combined with  Cartan's moving frame technique in terms of differential forms \cite{flanders} is a powerful tool also in diverse topics under the general heading of the  general theory of relativity. One of the main aims of the current paper is to show  that the linearization of Einstein's field equations provides another impressive illustrative case for the use of differential forms. 

The required mathematical tools are introduced in the following section in a sufficient generality  and the linearized quantities  are derived starting from scratch in any of the gravitational models studied. The collection of exterior algebra formulas introduced in the preliminary section pays off in the ensuing sections dealing with   Einstein's field equations along with the discussions of  the harmonic gauge condition and the quadratic Lagrangian 4-form from which the linearized equations follow. In Section 6, the Lagrangian and the corresponding field equations for a massive spin-2 field are discussed. Some further applications of the technique on the Newtonian limit, the definition of total energy for asymptotically flat isolated gravitating systems and the plane gravitational waves in the linearized approximation are briefly discussed in Section 7. As a relatively more involved application, a derivation of the  Lagrangian for new massive gravity in three dimensions is presented in some detail. In Section 8, a method to derive the linearized field equations for a given gravitational Lagrangian is developed and then applied to  some higher order gravitational models governed  by the Lagrangian densities $R^2$ and 
$g^{\mu\nu}\partial_\mu R\partial_\nu R$.  In sect. 9, the consequences of a nonvanishing torsion on the linearization  have been briefly discussed in the context of the Riemann-Cartan geometry. The paper ends with a final section with brief remarks concerning  Deser's construction of the Einstein-Hilbert Lagrangian.

\section{Mathematical preliminary}\label{preliminary}

This section provides a collection of exterior algebra formulas that are used extensively in the linearization technique, which can be found, for example,
in \cite{deser-isham,thirring,straumann} in sufficient detail.

\subsection{The basic rules of the exterior algebra}

The basic operators and the identities of the exterior algebra required manipulations and calculations are introduced in this section which also serves to fix the notation used. We denote the exterior product of forms by $\wdg$, whereas the exterior derivative by $d$, and the interior product with respect to a vector $X$ by $i_X$. These operators can be defined by their action on the exterior product of a $p$-form $\omega$ and a $q$-form  $\sigma$ as
\begin{eqnarray}
&d(\omega\wdg \sigma)
=
d\omega\wdg \sigma+(-1)^p\omega \wdg d\sigma,
\\
&i_X(\omega\wdg \sigma)
=
i_X\omega\wdg \sigma+(-1)^p\omega \wdg i_X\sigma.
\end{eqnarray}
Both operators are linear by definitions above and  $i_V$ is linear in its argument $V$ as well: $i_V=V^ai_a$ for a vector field $V=V^aX_a$.
Acting on a form, both operators are nilpotent $dd\equiv 0$  and $i_Vi_V\equiv0$. Two consecutive  interior derivatives with respect to vector fields $V$ and $W$ anti-commute $i_Vi_W=-i_Wi_V$. 

The set of an orthonormal basis 1-form will be denoted by $\{\theta^a\}$ whereas the  metric dual frame fields will be denoted by $\{X_a\}$.
The exterior product of basis 1-forms $\theta^a\wdg \theta^b\cdots $ will be abbreviated as $\theta^{ab\cdots}$. 
In terms of a natural coordinate basis, they can be expressed as $\theta^a=e^{a}_\mu dx^\mu$ and $X_a=e^{\mu}_{a}\partial_\mu$  where the matrices $e^{a}_\mu $ and $e^{\mu}_{a}$ are inverse of each other: $e_{\mu}^{a}e^{\nu}_{a}=\delta^{\nu}_{\mu}$, $e_{\mu}^{a}e^{\mu}_{b}=\delta^{a}_{b}$.
The interior product with respect to a basis vector field $X_a$ will be denoted by  $i_a\equiv i_{X_a}$ and note that
$i_a\theta^b=e_{a}^{\mu}e^{b}_{\nu}i_{\partial_\nu} dx^\mu=e_{a}^{\mu}e^{b}_{\mu}=\delta^{a}_{b}$. Since tensor components are used along with their components, the index raising/lowering of the tensor components have to be indicated in a slightly different manner. For example, the 1-form associated with a vector  
$V=V^{\mu}\partial_\mu=V^aX_a$ is denoted by $\tilde{V}\equiv V_\mu dx^\mu=V_a \theta^a$ with $V_\mu=g_{\mu\nu}V^\nu$. Likewise, the vector field 
associated with a 1-form $\sigma=\sigma_\mu dx^\mu$ is $\tilde{\sigma}=\sigma^\mu\partial_\mu$ with $\sigma^{\mu}=g^{\mu\nu}\sigma_\nu$.

The Hodge dual $*$ is a linear operator acting on the forms and it can be defined with a pseudo-Riemannian metric. 
The $*$ operator maps a $p$-form to a $(4-p)$-form. With the help of the $*$ operator, the inner product of two 1-forms provided by the metric can be extended to the definition of the inner product of two p-forms $\omega$ and $\sigma$: $\omega\wdg*\sigma=\sigma\wdg *\omega$. The action of the Hodge dual operator on basis $p$-forms can be defined with the help of a permutation symbol $\epsilon_{abcd}$ with $\epsilon_{0123}=+1$ while odd permutations of the indices resulting in $-1$, and vanishing otherwise. The invariant volume 4-form is defined as   $*1=\frac{1}{4!}\epsilon_{abcd}\theta^{abcd}$. The Hodge duals of basis 1-forms and 2-forms, for example, can be written as $*\theta^a=\frac{1}{3!}\epsilon^{a}_{\fant{a}bcd} \theta^{bcd}$, 
$*\theta^{ab}=\frac{1}{2}\epsilon^{ab}_{\fant{ab}cd} \theta^{cd}$. Acting on a $p$-form $**=(-1)^{p(n-p)+1}id$. Some identities involving $i_X, \wdg$ and $*$ which are of considerable use in exterior algebra manipulations are
\begin{eqnarray}
&i_X*\omega
=
*(\omega\wdg \tilde{X}),
\label{exterior-id1}\\
&\sigma \wdg *\omega
=
(-1)^{(p+1)}*i_{\tilde{\sigma}}\omega,\label{exterior-id2}
\end{eqnarray}
in terms of a $p$-form $\omega$, a 1-form $\sigma$ and a vector field $X$.
In terms of basis coframe and dual vector fields these read $i_a*\omega=*(\omega\wdg \theta_a)$ and $\theta^a\wdg *\omega=(-1)^{p+1}*i^a\omega$, respectively.

For the flat Minkowski spacetime, the Hodge dual will be denoted by $\star$ and the invariant volume element 
is $\star 1=dx^0\wdg dx^1\wdg dx^2\wdg dx^3$. The formulas for the orthonormal basis in a curved spacetime above hold for the natural basis of the Minkowski spacetime as well. The set of basis frame fields in the Minkowski spacetime is $\{\partial_a\}$ and the interior derivative with respect to the frame fields $\partial_a$ is  $i_a=i_{\partial_a}$. $\{\partial_a\}$ is dual to the set of natural basis $\{dx^b\}$, $i_adx^b=\delta^{a}_{b}$. Consequently, an expression of the form $i_adf$ is simply $\partial_a f$ for a 0-form $f$. For convenience of notation, the natural basis $dx^a$ which is the exterior derivative of the Cartesian coordinates $\{x^a\}$ will be denoted by $e^a$.

Finally, a cautionary remark related to the indices in the linearization is in order. The use of the  linearized tensorial quantities adopting an orthonormal coframe 
inevitably blurs the distinction between the indices relative to a coordinate basis labeled by Greek letters with those of  an orthonormal basis labeled by Latin letters. Both type of indices are raised and lowered by $\eta^{ab}$ and  $\eta_{ab}$, respectively.

\subsection{Cartan's structure equations}

The Levi-Civita connection 1-forms $\omega^{a}_{\fant{a}b}$ relative to an orthonormal basis satisfy Cartan's first structure equations 
\be\label{se1}
d\theta^a+\omega^{a}_{\fant{a}b}\wdg\theta^b=0,
\ee
satisfying  also the metric compatibility condition: $\omega_{ab}+\omega_{ba}=0$.
In terms of the Riemann tensor curvature 2-forms can be expressed in the form  $\Omega^{a}_{\fant{a}b}=\frac{1}{2}R^{ab}_{\fant{ab}cd}\theta^{cd}$, and in terms of the connection 1-forms, Cartan's second structure equations read
\be\label{se2}
\Omega^{a}_{\fant{a}b}
=
d\omega^{a}_{\fant{a}b}+\omega^{a}_{\fant{a}c}\wdg\omega^{c}_{\fant{a}b}.
\ee
The curvature 2-forms satisfy $\Omega_{ab}+\Omega_{ba}=0$ and the first and second Bianchi identities, 
namely, $\Omega^{a}_{\fant{a}b}\wdg \theta^b=0$ and $D\Omega^{ab}=0$, respectively. 

$D$ stands for the covariant exterior derivative acting on 
tensor valued-forms \cite{straumann,thirring,benn-tucker}. For example, acting on vector components $V^a$, it yields $DV^a=dV^a+\omega^{a}_{\fant{a}b}V^b$
whereas, the second Bianchi identity explicitly reads 
$D\Omega^{a}_{\fant{a}b}=d\Omega^{a}_{\fant{a}b}+\omega^{a}_{\fant{a}c}\wdg \Omega^{c}_{\fant{a}b}-\omega^{c}_{\fant{a}b}\wdg \Omega^{a}_{\fant{a}c}$.
The linearization procedure  will often turn a covariant derivative into an exterior derivative and the covariant exterior derivative will not play a significant role in the linearized equations. 

In a typical curvature calculation,  one first solves (\ref{se1}) in terms of a given orthonormal basis and then inserts them into (\ref{se2}) as will be
repeated below in a general form at the linearized level. As a typical calculation in exterior algebra, one can verify that the first structure equations can uniquely be solved for the connection 1-forms \cite{thirring};
\be\label{inverted-se1}
\omega^{a}_{\fant{b}b}
=
\frac{1}{2}i^ai_b (d\theta^c\wdg \theta_c)-i^ad\theta_b+i_bd\theta^a
\ee
by calculating two successive inner products of (\ref{se1}) with respect to basis vector fields.

Finally, note that the Ricci 1-forms are defined in terms of the Ricci tensor $R^a\equiv R^{a}_{\fant{a}b}\theta^b=R^{ac}_{\fant{ac}bc}\theta^b$ which can also be given by $R_a=i^b\Omega_{ba}$ and thus the Ricci scalar is $R=i_aR^a$. Likewise, Einstein 1-forms can be written as $G^{a}\equiv R^a-\frac{1}{2}R\theta^a$.

\subsection{Metric perturbation 1-forms}

Weak gravitational fields, considered as  small perturbations to the flat Minkowski space are studied 
by introducing metric coefficients $h_{ab}(x^c)$ which are assumed to satisfy $|h_{ab}|\ll 1$ for consistency.
In a mathematically precise manner, a weak gravitational field has a metric $g=\eta_{ab}\theta^a\ot \theta^b$ of the approximate form
\be\label{lin-metric}
g^L= \eta+2h_{ab}dx^{a}\ot dx^{b}=\eta+h_{a}\ot e^{a}+e^{a}\ot h_{a}.
\ee
The flat spacetime metric is of the form  $\eta=\eta_{ab}e^a\ot e^b$ where $\eta_{ab}=diagonal(-+++)$. 
In order to implement the exterior algebra of forms in a flat background, it is convenient to introduce the 1-forms $h^{a}=h^{a}_{\fant{a}b}e^b$ instead of the symmetric scalar components $h_{ab}=h_{ba}$. Note that (\ref{lin-metric}) implies that the symmetry of  $h_{ab}$ is related with the symmetry  $g^L_{ab}=g^L_{ba}$.
To denote the trace of the perturbation coefficients $\eta_{ab}h^{ab}=h^{a}_{\fant{a}a}\equiv h$ will be used. 
To first order in $h_{ab}$, a total basis 1-forms $\theta^a$ then decomposes as
\be\label{lin-basis-coframe}
\theta^a_L
=
e^a+h^a
\ee
as a 1-form equation. The label $L$  in the expression, and in what follows, refers to a quantity linear in $h_{ab}$.
For the  equations that are valid in the linear approximation, the equality sign will be used as has been done in (\ref{lin-metric}) and (\ref{lin-basis-coframe}), the $O(h^2)$-terms are assumed to be ignored. The linearized forms of the basis $p$-forms and their Hodge duals can simply be found by making use of (\ref{lin-basis-coframe}). For example, one can readily obtain the following linear approximations:
\begin{eqnarray}
\left(\theta^{a}\wdg \theta^{b}\right)_L
&=&
e^{a}\wdg e^b+e^a\wdg h^b+h^a\wdg e^b,
\\
\left(\theta^{a}\wdg \theta^{b}\wdg \theta^c\right)_L
&=&
e^{a}\wdg e^b \wdg e^{c}+e^a\wdg e^b\wdg h^c+e^a\wdg h^b\wdg e^c
\nonumber\\
&&\qquad+h^a\wdg e^b\wdg e^c,
\\
\left(\theta^{a}\wdg \theta^{b}\wdg \theta^c\wdg \theta^d\right)_L
&=&
e^{a}\wdg e^b \wdg e^{c}\wdg e^d
+
e^a\wdg e^b\wdg e^c\wdg h^d
+
e^a\wdg e^b\wdg h^c\wdg e^d
\nonumber\\
&&\qquad +e^a\wdg h^b\wdg e^c\wdg e^d+h^a\wdg e^b\wdg e^c\wdg e^d,
\label{basis_expansion}\\
\left[*\theta^{a}\right]_L
&=&
\frac{1}{6}\epsilon^{a}_{\fant{a}bcd}\left(\theta^{b}\wdg \theta^{c}\wdg \theta^{d}\right)_L
=
\star e^{a}+h_b\wdg \star e^{ab},
\\
\left[*(\theta^{a}\wdg \theta^{b})\right]_L
&=&
\frac{1}{2}\epsilon^{ab}_{\fant{ab}cd}\left(\theta^{c}\wdg \theta^{d}\right)_L
=
\star e^{ab}+h_c\wdg \star e^{abc},
\\
\left[*(\theta^{a}\wdg \theta^{b}\wdg \theta^c)\right]_L
&=&
\epsilon^{abc}_{\fant{abc}d}\theta^{d}_L
=
\star e^{abc}+h_d\epsilon^{abcd},\label{hodge_basis_expansion}
\end{eqnarray}
where it is assumed that terms higher then $O(h)$ are omitted from the expressions on the right hand sides. The linearized forms of the basis $p$-forms and their Hodge dual can likewise be found by using expansions akin to (\ref{lin-basis-coframe}) for  given tensorial expressions. One can also find
$(*1)^L=*^L1=(1+h)\star1$ using the above formulas.

Finally, note that one can linearize  the frame fields as $X_a^L=\partial_a-h_{ab}\partial^{b}$ and consequently, 
$(i_{X_a}\theta^b)^L=i_{X^L_a}\theta^{a}_L=\delta^{b}_{a}$ to the \emph{first} order in $h$.
Accordingly, the interior product can be linearized as $(i_{X_a})^L=i_{X^L_a}=i_{(\partial_a-h_{ab}\partial^b)}=i_{\partial_a}-h_{ab}i_{\partial^b}$. The linerization formulas of this section  are sufficient to rewrite the linearized forms of identities (\ref{exterior-id1}) and (\ref{exterior-id2}) provided
in the previous section.

\subsection{The linearized connection and curvature forms}

The above formulas are sufficient to introduce the first order approximation into the first structure equations (\ref{se1}).
One can show that they become
\be\label{lin-se1}
dh_a+\omega^{L}_{ab}\wdg e^b=0.
\ee
The linearized Levi-Civita connection is assumed to retain the metric compatibility and therefore one has $\omega^L_{ab}+\omega^L_{ba}=0$.
As for the structure equations, the linearized structure equations (\ref{lin-se1}) can be inverted to find $\omega^L_{ab}$, for example, by making use of the linearized version of the formula  (\ref{inverted-se1}). One finds
\be\label{lin-conn-1-form}
\omega^{L}_{ab}
=
-i_adh_b+i_bdh_a.
\ee
Note that the components of the connection 1-forms can be read off from the expression
\be\label{connection-form2}
\omega^{L}_{ab}
=
-(\partial_a h_{bc}-\partial_b h_{ac})e^c.
\ee
Consequently, the linearized curvature 2-forms can be written as
\be\label{lin-curv-2-form}
\Omega^{L}_{ab}
=
d\omega^{L}_{ab}
\ee
by dropping $\omega^2$ terms in (\ref{se2}). The components of the linearized Riemann tensor, $R^{ab}_{Lcd}$ can be found by inserting (\ref{lin-conn-1-form}) into the linearized second structure equations and noting that $\Omega^{ab}_L=\frac{1}{2}R^{ab}_{Lcd}\left(\theta^{c}\wdg \theta^d\right)_L$.
One finds
\be\label{lin-riem-comp}
R_{abcd}^L
=
\partial_a \partial_d h_{bc}-\partial_b\partial_d h_{ac}
-
\partial_a \partial_c h_{bd}+\partial_b\partial_c h_{ad}.
\ee

In contrast to the expressions  (\ref{connection-form2}) and 
(\ref{lin-riem-comp}), the use of the expressions in differential forms, namely (\ref{lin-conn-1-form}) and (\ref{lin-curv-2-form}) is more convenient as will be justified below.

\section{The linearized Einstein tensor}

Einstein's field equations can be expressed in various equivalent forms. The form suitable for the current presentation involves the Einstein 1-forms 
$G^a=G^{a}_{\fant{b}b}\theta^b$ \cite{thirring,straumann}. The Hodge dual of the Einstein 1-forms can be conveniently expressed in the form
\be\label{einstein-3-form-def}
*G^a
=
-\frac{1}{2}\Omega_{bc}\wdg *\theta^{abc}.
\ee
The expression on the right hand side can readily be obtained by a coframe variation of the Einstein-Hilbert Lagrangian expressed in terms of 
curvature 2-forms \cite{thirring,straumann,benn-tucker}. One can also verify that the expression (\ref{einstein-3-form-def}) indeed 
leads to the Einstein tensor by first rewriting it in the form 
$*G^a=-\frac{1}{4}R^{bc}_{\fant{bc}mn}\epsilon^{a}_{\fant{a}bcd}\theta^{mnd}.$
Taking the Hodge dual, one finds
\be\label{einstein-1form4}
G^a
=
\frac{1}{4}\epsilon^{abcd}\epsilon_{kmnd}R_{bc}^{\fant{bc}mn}\theta^k.
\ee
Now, by making use of the identity involving the product of permutation symbols $\epsilon^{abcd}\epsilon_{kmnd}=\delta^{abc}_{kmn}$
where $\delta^{abc}_{def}$ is the generalized Kronecker symbol, (\ref{einstein-1form4}) can be rewritten in the form  
\be
G^a
=
\frac{1}{4}\delta^{abc}_{kmn}R^{mn}_{\fant{mn}bc}\theta^k.
\ee
The familiar expression $G^{a}\equiv G^{a}_{\fant{a}b}\theta^b=(R^{a}_{\fant{a}b}-\frac{1}{2}\delta^{a}_{b}R)\theta^b$ now can be obtained  
by expanding the generalized Kronecker symbol into a product of  Kronecker deltas as
\be\qquad
\delta^{adf}_{bce}
=
\delta^{a}_{b}\delta^{df}_{ce}
-
\delta^{d}_{c}\delta^{af}_{be}
+
\delta^{f}_{e}\delta^{ad}_{bc}\qquad \mbox{and}\qquad \delta^{ab}_{cd}=\delta^{a}_{c}\delta^{b}_{d}-\delta^{a}_{d}\delta^{b}_{c}.
\ee

As  is now verified, the linearized Einstein tensor can be found by linearizing the expression (\ref{einstein-3-form-def}). For convenience, one can 
define $\star G^a_L\equiv (*G^a)_L$, where the linearized Einstein 1-forms then have the expansion $G_a^L\equiv G^L_{ab}e^b$. Explicitly, one has
\be
\star G^a_L
=
-\frac{1}{2}\Omega^L_{bc}\wdg \star e^{abc}
=
-\frac{1}{2}d\omega^L_{bc}\wdg \star e^{abc}.
\ee
The factor $\Omega^L_{bc}$ survives the exterior multiplication since it has no $O(h)$ terms and  in addition, because $e^a$ is a natural (\emph{Cartesian}) basis, 
$d\star e^{abc}=0$ identically as a consequence of the identity  for exterior derivative: $dd\equiv0$. Consequently, the linearized Einstein 3-forms take the form
\be\label{lin-ein-f1}
\star G^a_L
=
-\frac{1}{2}d\left(\omega^L_{bc}\wdg \star e^{abc}\right).
\ee
Note that the linearized vacuum equations take the remarkable form $d\star F^a_L=0$ having a formal resemblance  to  Maxwell's equation $d*F=0$, where $F$ is the Faraday 2-form, in terms of the Thirring 2-forms $F^a$ which are defined \cite{thirring}  as $*F^a=-\frac{1}{2}\omega_{bc}\wdg*\theta^{abc}$.

By making use of the expression (\ref{lin-conn-1-form}) for the  connection 1-forms, (\ref{lin-ein-f1}) reduces to
\be\label{lin-ein-3-form}
\star G^a_L
=
d\left(i_bdh_c \wdg \star e^{abc}\right).
\ee
Further simplification of the resulting expression can be achieved by noting that
\be
i_adh_b
=
i_a(dh_{bc}\wdg e^c)
=
(i_adh_{bc})e^c
-
dh_{ba}.
\ee
By making use of the symmetry property, $h_{ab}=h_{ba}$, one finally ends up with
\be\label{lin-ein-form2}
\star G^a_L
=
d\left[e^b\wdg \star (dh_b\wdg e^{a})\right].
\ee
This concise  expression for the Einstein 3-forms can be presented in a variety of alternative forms. For example, it can be rewritten as 
\be\label{eins-form-form3}
\star G^a_L
=
d \star dh^a +d \star \left[(i_bdh^b)\wdg e^{a}\right].
\ee
An immediate observation about the expression (\ref{lin-ein-form2}) is that it implies that the linearized Bianchi identity $(D*G^a)_L=0$ amounts to the identity $dd\equiv0$. This also directly follows from the linearized Bianchi identity $(D\Omega^{ab})^L=d\Omega^{ab}_{L}\equiv 0$.

The components of the linearized Einstein tensor  can now be derived by noting that $G^{ab}_L\star1=e^a\wdg \star G^b_L$ and by making use of the identity 
$e_{abc}\wdg\star e^{def}=\delta^{abc}_{def}\star1$, so that (\ref{lin-ein-form2}) eventually leads to 
\be\label{einstein-comp-1}
G^a_{Lb}
=
\delta^{adf}_{bce}\partial^{c}\partial_{d}h^{e}_{\fant{i}f}.
\ee
The result expressed in (\ref{einstein-comp-1}) can be expanded into the familiar expression
\be\label{lin-einstein-ten}
G^L_{ab}
=
-
\qed h_{ab}
-
\eta_{ab}\partial_{c}\partial_{d}h^{cd}
+
\partial_{a}\partial^ch_{cb}
+
\partial_{b}\partial^ch_{ca}
-
\partial_{a}\partial_bh
-
\eta_{ab}\qed h
\ee
where $\qed\equiv \eta^{ab}\partial_a\partial_b$.
The expression on the right hand side of (\ref{lin-einstein-ten}) can be obtained by expanding the generalized Kronecker symbol into a product of the Kronecker deltas. The explicit expression on the right hand side shows that the expression in (\ref{lin-ein-3-form}) leads to a symmetric linearized Einstein tensor.
One can verify that the expression (\ref{lin-einstein-ten}) follows from (\ref{lin-riem-comp})
by using $G^L_{ab}=R^L_{ab}-\frac{1}{2}\eta_{ab}R^L$ with  $R^L_{ac}=\eta^{bd}R^L_{abcd}$ and $R^L=\eta_{ab}R^{ab}_L$.
The particular staggering of partial derivative indices in the expression (\ref{lin-einstein-ten}) is reproduced by  exterior algebra from the compact expression 
(\ref{lin-ein-form2}). It is worth to emphasize that the  expression (\ref{lin-einstein-ten}) is the expression obtained by making use of the linearized Christoffel 
symbols and the linearized Riemann tensor and that (\ref{lin-einstein-ten}) is merely the expression of (\ref{lin-ein-form2}) in component form. 

Finally, the linearized scalar curvature $R^L$ can simply be  obtained from (\ref{lin-einstein-ten}), or else by using the definition 
$R^L\star1=-e_a\wdg \star G^a_L$ and (\ref{lin-ein-form2}). One finds
\be\label{lin-scalar-curvature}
R_L\star1
=
2d\star i_bdh^b
=
2(\partial_a\partial_b h^{ab}-\qed h)\star1.
\ee
Using the above results, one can show that the Hodge duals of the Ricci 1-forms are given by
\be\label{Ricci-1-form1}
\star R^a_L=d\star dh^a+L^a\star i_bdh^b
\ee
with $L_a\equiv i_ad+di_a$  standing for the Lie derivative  with respect to $\partial_a$ on a $p$-form.
As a consistency check, one can calculate the components of the linearized Ricci 1-forms by using $R_a^L=i^b_L\Omega^{L}_{ba}$ with those of obtained from 
(\ref{Ricci-1-form1}) as well.

\section{Gauge fixing: The harmonic gauge}

For harmonic coordinates, a gauge condition, also called harmonic gauge condition, can be  expressed as \cite{thirring}
\be\label{harmonic-gauge-cond}
d\star dx^a=0.
\ee
On the other hand, by making use of $e^a\approx\theta^a-h^a$ and the identity $d*\theta^a=-\omega^{a}_{\fant{a}b}\wdg*\theta^b$, one finds
\be\label{gauge-con1}
d\star h_a+\omega^L_{ab}\wdg \star e^b=0
\ee
to first order in $h_{ab}$.
The expression (\ref{gauge-con1}) leads to the following identity on the derivatives of the perturbation 1-forms
\be\label{gauge-cond-form1}
i_adh^a+\frac{1}{2}dh=0.
\ee
In the familiar component form, (\ref{gauge-cond-form1}) can be rewritten as
\be\label{gc1}
\partial_a h^{a}_{\fant{a}b}-\frac{1}{2}\partial_bh=0.
\ee
With the gauge  condition imposed, the linearized Einstein 3-forms become
\be\label{gf-einstein-form}
\star G^a_L
=
d\star d \bar{h}^a
\ee
where  new perturbation 1-forms $\bar{h}^a$ are defined in terms of the original as
\be\label{gauge-fix}
\bar{h}_a
=
\left(
h_{ab}-\frac{1}{2}\eta_{ab}h
\right)e^b.
\ee
In terms of $\bar{h}^a$, the gauge condition (\ref{gc1}) simply reads $\partial_a \bar{h}^{a}_{\fant{a}b}=0$. Consequently, the linearized Einstein 
tensor reduces to the form
\be\label{gauge-cond2}
G_{ab}^L
=
-\qed \bar{h}_{ab}
\ee
in the harmonic gauge. In the component form of (\ref{gauge-cond2}), the gauge condition $\partial^a \bar{h}_{ab}=0$ implies the linearized Bianchi identity $\partial^a G_{ab}^L=0$  as a consequence of the fact that $\partial_a$ commutes with $\qed$.

\section{The Pauli-Fierz Lagrangian}

Einstein's vacuum field equations follow from the Einstein-Hilbert Lagrangian 4-form 
\be\label{EH-lag}
L_{EH}
=
\frac{1}{2}R*1
=
\frac{1}{2}\Omega_{ab}\wdg *\theta^{ab}.
\ee
Equation (\ref{EH-lag}) is a special Lagrangian in the sense that it leads to second order partial differential equations in the metric components despite the fact that it
contains second order derivatives. On the other hand, the parts containing the second order derivatives can be relegated to a boundary term  by making use of the identity
\be
d*\theta^{ab}
=
-
\omega^{a}_{\fant{a}c}\wdg *\theta^{cb}
-
\omega^{b}_{\fant{a}c}\wdg *\theta^{ac}
\ee
(that follows from $D*\theta^{ab}=0$ satisfied by a Levi-Civita connection) in conjunction with the second structure equations in (\ref{EH-lag}). 
One finds
\be\label{EH-lag2}
L_{EH}
=
-
\frac{1}{2}\omega_{ac}\wdg \omega^{c}_{\fant{a}b}\wdg *\theta^{ab}
-
\frac{1}{2}d\left(\omega_{ab}\wdg*\theta^{ab}\right)
\ee
The field equations which  are first order in the metric perturbations require a Lagrangian which is second order in the metric perturbations
and (\ref{EH-lag2}) is evidently in a suitable form to obtain such a Lagrangian:
\be\label{reduced-EH-lag1}
L^{(2)}_{EH}[h]
\equiv
-
\frac{1}{2}\omega^L_{ac}\wdg \omega^{c}_{Lb}\wdg \left(*\theta^{ab}\right)_L.
\ee
Now retaining the terms to order $O(h^2)$ in this expression, one ends up with
\be\label{red-EH-lag}
L^{(2)}_{EH}[h]
=
-\frac{1}{2}e_a\wdg dh_b\wdg\star\left(e^b\wdg dh^a\right),
\ee
after some exterior algebra calculations using the formulas given in Section \ref{preliminary}. The Lagrangian $L^{(2)}[h]$ can be regarded as a Lagrangian for  the 1-form field $h_a$ in Minkowski spacetime. A Lagrangian density involving the partial derivatives  can be explicitly obtained from  
(\ref{red-EH-lag}). One finds
\be\label{red-EH-lag2}
L^{(2)}_{EH}[h]
=
\left(
-
\frac{1}{2}\partial_{c}h^{ab}\partial^{c} h_{ab}
+
\partial_{a}h^{ab}\partial_{b} h
-
\partial_{c}h_{ab}\partial^{b} h^{ca}
+
\frac{1}{2}\partial_{a}h\partial^{a} h
\right)\star1
\ee
up to an omitted total derivative. Equation (\ref{red-EH-lag2}) is nothing but the Pauli-Fierz Lagrangian density.

The form of the Lagrangian 4-form (\ref{red-EH-lag}) suggests  still another way to derive the linearized Einstein 3-forms, or equivalently, the linearized Einstein tensor. More precisely, one can work out the variational derivative $\delta L^{(2)}[h]\equiv L^{(2)}[h^a+\delta h^a]-L^{(2)}[h^a]$ with respect 
to the perturbation 1-forms $h^a$ to obtain 
\be\label{h-var}
\delta L^{(2)}_{EH}[h]
=
-
\delta h_a\wdg d \left[e^b\wdg\star\left(dh_b\wdg e^a\right)\right],
\ee 
up to an omitted boundary term and holding the basis 1-forms $e^a$ of the Minkowski spacetime fixed.
Consequently, by a comparison of the result (\ref{h-var}) with (\ref{lin-ein-form2}), one can deduce 
$
{\delta L^{(2)}_{EH}}/{\delta h_a}
=
-\star G^a_L
$
which has the exact variational derivative counterpart
$
{\delta L_{EH}}/{\delta \theta_a}
=
-* G^a
$ (See, for example, the derivation provided in \cite{thirring,straumann}).

Finally, note that the fixing of the harmonic gauge can also be achieved by supplementing the Pauli-Fierz Lagrangian with a term $L_{gf}$ of the form
\be
L_{gf}
=
\frac{1}{2}(i_adh^a+dh)\wdg\star(i_bdh^b+dh)
\ee
involving the square of the gauge condition. By simplifying the  expression on the right hand side, one can obtain an extended Lagrangian of the form
\be\label{gf-PF-lag}
L'^{(2)}_{EH}
\equiv
L^{(2)}_{EH}
+
L_{gf}
=
-
\frac{1}{2}dh_a\wdg \star dh^a
+
\frac{1}{2}dh\wdg e^a\wdg\star dh_a
+
\frac{1}{8}
dh\wdg\star dh.
\ee
One can rederive the field equations (\ref{gf-einstein-form}) and the harmonic gauge condition (\ref{gauge-cond-form1}) by calculating the  variational derivative of the extended Lagrangian (\ref{gf-PF-lag}) with respect to the variables $h_a$ and $h$, respectively.

Compared to the coordinate methods making use of the scalar functions $h_{ab}$, the linearization technique developed above in terms of 1-forms $h_a=h_{ab}e^b$ provides a practical and powerful calculational technique, often rendering the equations more transparent as well. This claim can be justified by the illustrative examples below.

\section{The Lagrangian for a massive spin-2 field in vacuum}

The field equations for a massive spin-2 field was introduced almost 80 years ago by  Pauli and Fierz \cite{fierz-pauli1,fierz-pauli2}. The issue of massive spin-2 modes for the gravitational interactions in a cosmological context frequently occurs in the modified gravity models, for example, to explain the cosmic  accelerated expansion of the Universe (See, for example, \cite{deser-can-j-phys,deser-waldron-zahariade,hinterbichler,derham} for a general discussion on the graviton mass). In three dimensions,  massive gravity models, such as New massive gravity to be discussed briefly below,  are important since they provide  toy models in the context of quantum theory of gravity.

The introduction of mass to a spin-2 field has some interesting and subtle features requiring some extra attention even at the linearized level in vacuum.
Mathematically admissible terms that one can consider for the massive spin-2 field are of the form $h^2$, or $h^{ab}h_{ab}$ or else some particular combination of these quadratic terms. Let us consider, for example, the Lagrangian 4-form 
\be\label{mGR-lag}
L^{(2)}_m[h]
=
L^{(2)}_{EH}[h]
+
\frac{m^2}{2}e^a\wdg h_b\wdg\star\left(e^b\wdg h_a\right)
\ee
where the mass term can be rewritten in the form
\be
e^a\wdg h_b\wdg\star\left(e^b\wdg h_a\right)
=
\left(h_{ab}h^{ab}-h^2\right)\star 1.
\ee
The significance of the particular mass term will be apparent as one proceeds with the derivation of  the field equations  from the Lagrangian 4-form (\ref{mGR-lag}).

The linearized coordinate covariance expressed in terms of a vector field $X=\xi^a\partial_a$ as $x^a\mapsto x'^a= x^a+\xi^a(x)$ leads to  
$h_{ab}\mapsto h'_{ab}=h_{ab}+\partial_a\xi_b+\partial_b\xi_a$. The infinitesimal change in the metric perturbation 1-forms can be expressed in the form language as 
$\delta h_a=di_a\tilde{X}+i_ad\tilde{X}\equiv L_a\tilde{X}$ 
with $\tilde{X}=\xi_a e^a$. One can show that $\Omega^{ab}_L$ remains invariant under these transformations. Consequently, the ``Bianchi identity" for the massive gravity Lagrangian (\ref{mGR-lag}) that follows from the linearized coordinate covariance can be obtained by inserting $\delta h_a=L_a\tilde{X}$ to the variational derivative:
\be\label{var-mGR-lag}
\delta L^{(2)}_m[h]
=
\delta L^{(2)}_{EH}[h]-m^2\delta h_a \wdg e^b\wdg\star\left(e^a\wdg h_b\right)=0.
\ee
Using the fact that $\delta L^{(2)}_{EH}[h]$ leads to an exact 3-form (linearized Bianchi identity for the massless case), (\ref{var-mGR-lag}) then leads to
\be
d\left[e^a\wdg \star\left(h_a\wdg e^b\right)\right]=0
\ee
which may be rewritten conveniently as 
\be\label{id1-massive}
d\star\left(h^a-he^a\right)=0.
\ee
In the component form, this identity can be rewritten as $\partial_ah^a_{\fant{a}b}-\partial_bh=0$,
or equivalently, in a concise form as $i_a dh^a=0$. The latter form of the identity also offers some  insight 
into the massive spin-2 field equations: This condition cancels out the scalar curvature term  from the field equations recalling that 
the linearized scalar curvature can be written in the form $R^L=2\star d\star i_adh^a$.

The identity (\ref{id1-massive}) then can be used to simplify the  massive gravity equations. Explicitly, the massive spin-2 equations read
\be\label{mGR-eqn}
d\left[e^b\star \left(d h_b\wdg e^a\right)\right]
-
m^2 e^b\wdg\star \left(e^a\wdg h_b\right)=0.
\ee
The trace of the field equations can be calculated by wedging (\ref{mGR-eqn}) with $e_a$ from the left.The trace of the first term leads to the scalar curvature, which vanishes by the ``Bianchi identity"' and one can verify that, in vacuum, the trace of the mass term reads $h=0$. Hence the Pauli-Fierz mass term in 
(\ref{mGR-lag}) is a unique quadratic combination  of the perturbation 1-forms that leads to such a simplification.
Taking the simplifications into account, and taking the Hodge dual of (\ref{mGR-eqn}), the field equations for a massive spin-2 field eventually reduce to the form
\be
(\star d\star d+ m^2)h_a=0
\ee
as a 1-form equation.
In components, this can be rewritten as
\be
(\qed-m^2)h_{ab}=0
\ee
and these are compatible with the subsidiary conditions $\partial^ah_{ab}=0$ and $h=0$ by construction. Equation (\ref{id1-massive}) is usually referred to as  a constraint for the massive spin-2 field $h_{ab}$ because the constraints are not gauge invariant \cite{boulware-deser}. 
The reason for this is that as a consequence of these constraints  $R^L$ vanishes as one can observe from Eq. (\ref{lin-scalar-curvature}). However, $R^L$ is invariant under  the gauge transformations $\delta h_{ab}=\partial _a\xi_{b}+\partial_b\xi_{a}$ generated by a vector field $X=\xi^a\partial_a$ and it cannot be set to zero by a gauge choice.

\section{Some applications}

\subsection{Newtonian Limit}

In order to recover the Newtonian limit of the field equations, let us consider a static metric of the form 
\be\label{bt-metric}
g=-F^2dt\ot dt+F^{-2}\delta_{ij}dx^i\ot dx^j
\ee
with a metric function $F=F(x^a)$ having the approximate form $F^2\approx 1+2h(x^a)$. In this approximation, only the diagonal components of the metric perturbations survive and one has $\theta^0_L=e^0+he^0$ and $\theta^i_L=e^i-he^i$ for $i=1, 2, 3$. The linearized Einstein's equations,
$\star G^a_L=\kappa \star T^a$, follow from the total Lagrangian density $L_{tot}=L^{(2)}[h]+L_{int.}$, 
with the matter-interaction term $L_{int}\equiv \kappa h_a\wdg\star T^a$
and the coupling constant $\kappa$. An ideal fluid energy-momentum tensor $T\equiv T_{ab}\theta^a\ot \theta^b$ is of the form 
$T=(\rho+p)\tilde{U}\ot \tilde{U}+pg$ in terms of the proper energy density $\rho(x^i)$, pressure $p$ and four-velocity $U=U^\mu\partial_\mu$. For $p=0$ and for 
$a=0$, the field equations lead to the Poisson equation $\star d\star  dh=-4\pi G \rho$ in three space dimensions where one can identify $h$ as the Newtonian gravitational potential and the coupling constant $\kappa$ in the Einstein field equations can be expressed  in the form $\kappa^{-1}=16\pi G /c^4$, in terms of the Newtonian gravitational constant $G$ and the speed of light $c$. In addition, one can show  that the geodesics of  massive test particles in the metric (\ref{bt-metric}) lead to
$\ddot{x}^i+\frac{1}{2}\partial_i F^2\approx 0$ assuming that the slow-motion approximation  $|\dot{x}^i|\ll 1$ is valid in addition to the weak-field approximation  $|h_{ab}|\ll1$. Here a dot over a spatial coordinate refers to the derivative with respect to  proper time $\tau$, $\dot{x}^i={dx^i}/{d\tau}$
\cite{benn-tucker}.

\subsection{Total energy of an asymptotically flat system} 

For an asymptotically flat spacetime of an isolated gravitating system, the total energy defined by Arnowitt-Deser-Misner (ADM) has the explicit coordinate expression of the form
\be\label{ADM-original-def}
M_{ADM}
=
\frac{1}{16\pi G}\int_{S_{\infty}}dS^i \left(\partial_k g_{ki}-\partial_{i}g_{kk}\right)
\ee
where $dS^i$ is the area element of a two-dimensional surface \cite{abbott-deser}. 
It is possible to recover (\ref{ADM-original-def}) by using  the Thirring 2-forms provided that a coordinate basis  is adopted in the expression (\ref{ADM-thirring-def}) below \cite{thirring}. By taking  the decomposition of the Einstein 3-forms of the form $*G^\mu=d*F^\mu+*t^\mu$ into account \cite{thirring,straumann} and using Stokes' theorem \cite{flanders} to convert the integral over three-dimensional space into a flux integral, one obtains  a four-vector $P^\mu$ defined by 
\be\label{ADM-thirring-def}
P^\mu
=
-\frac{1}{16\pi G}\int_{S_\infty}*F^\mu.
\ee
$M_{ADM}$ corresponding to the temporal component of the four vector (\ref{ADM-thirring-def}) can be calculated by integrating the exact 3-form part of  $*G^0$ over a 2-sphere of infinite radius  \cite{thirring,straumann}. For instance, one can show that $M_{ADM}$ is the Schwarzchild mass by making use of 
(\ref{ADM-thirring-def}).
However, (\ref{ADM-original-def}) can be recovered at the linearized level as well. Assuming that the  metric for a given static spacetime can be brought to the form (\ref{lin-metric}), one can use the linearized form of the Thirring 2-forms, namely $\star F^a_L\equiv(-\frac{1}{2}\omega_{bc}\wdg *\theta^{abc})_L$
with $F_a^L\equiv\frac{1}{2}F^{L}_{abc}e^{b}\wdg e^c$, to obtain
\be
P^0
=
-\frac{1}{16\pi G}\int_{S_\infty}\star F^0_L
=
-\frac{1}{16\pi G}\int_{S_\infty} \left(\partial_k h^{k}_{\fant{k}i}-\partial_i h^{k}_{\fant{k}k}\right)\star e^{i0}
\ee
with $i,k=1, 2, 3$ and the integration measure  is now built into the expression which is to be restricted to the sphere of radius $R$ and a limiting procedure $R\mapsto \infty$, keeping $x^0=ct$ constant.

The Abbot-Deser-Tekin  charges \cite{abbott-deser} are a generalization of the ADM energy defined on asymptotically curved backgrounds and applied to the models derived from general quadratic curvature  gravitational Lagrangians \cite{deser-tekin3}. The energy definition provided by  Deser and Tekin \cite{deser-tekin2} is then reformulated in terms of differential forms in \cite{baykal} using the mathematical formalism as presented above.

\subsection{Gravitational Waves}

 A plane gravitational wave solution, propagating along the $x^3$-axis, can be constructed by inserting the
plane wave metric ansatz into the vacuum field equations $\qed \bar{h}_{ab}=0$. The ansatz  involves a null propagation vector of the form
$k=e^0\mp e^3$ and a constant polarization tensor $\epsilon_{ab}$ so that  perturbation metric $h_{ab}$ is in the form  corresponding to  the real part of 
$\epsilon_{ab}e^{ik_cx^c}$. By imposing the so-called transverse-traceless gauge conditions ($\partial^a h_{ab}=h=0$),  $+$ and $\times$ transverse polarization modes for the plane gravitational waves are then obtained in terms of the nonvanishing  metric perturbations $h_{12}=h_{21}$ and $h_{22}=-h_{11}$ respectively (See, for example, the lucid presentation in \cite{straumann}).

More generally, it is well known that the plane-fronted gravitational waves with parallel-propagation ($pp$-waves) constitute their own linearization. The Kerr-Schild form of the metric of the $pp$-waves can be written explicitly as \cite{exact-sol}
\be\label{pp-wave-metric}
g=
\eta-2H du\ot du
\ee
in terms of the global null coordinates $x^a=\{u, v, \zeta, \bar{\zeta}\}$ where $u,v$ are the real null coordinates related with the Cartesian coordinates as 
\be
u=
\frac{1}{\sqrt{2}}\left(x^0-x^3\right),\qquad 
v=
\frac{1}{\sqrt{2}}\left(x^0+x^3\right),
\ee
whereas the complex conjugate coordinate $\zeta$ and $\bar{\zeta}$ are related to the Cartesian as
\be
\zeta
=
\frac{1}{\sqrt{2}}(x^1+ix^2),
\qquad
\bar{\zeta}
=
\frac{1}{\sqrt{2}}(x^1-ix^2).
\ee
In terms of the null coordinates, the Minkowski spacetime  metric explicitly reads
\be\label{flat-null-metric}
\eta=
-du\ot dv-dv\ot du+d\zeta\ot d\bar{\zeta}+d\bar{\zeta}\ot d\zeta
\ee
The nonvanishing metric coefficients in this case are $-\eta_{01}=-\eta_{10}=\eta_{12}=\eta_{21}=1$.
The coordinates of the null coframe are related with the Cartesian coordinates by a transformation matrix with
constant coefficients and therefore the (linearized) field equations relative to a null coframe retain the form (\ref{lin-ein-3-form}). However, the indices relative to the null coframe  in the null coframe case are  lowered and raised by the $\eta_{ab}$ and $\eta^{ab}$ of the metric (\ref{flat-null-metric}), respectively.

For the $pp$-wave metric ansatz (\ref{pp-wave-metric}), the real profile function $H=H(u,\zeta,\bar{\zeta})$ corresponds to the only nonvanishing metric perturbation 1-form $h_{0}=-h^1=-Hdu$, where the numerical indices refer to the null coframe. An explicit expression for  the only nonvanishing Einstein 3-form can be calculated  simply by making use of (\ref{lin-ein-3-form}) as
\be
\star G^1
=
e^0\wdg d\star\left(dh_0\wdg e^1\right)
=
-e^0\wdg d\star\left(dH\wdg e^{01}\right)
\ee
where $e^0=du, e^1=dv$ and $e^2=d\zeta=\bar{e}^3$. One can show that the expression on the right hand side reduces  to
\be\label{eqn-for-profile-funct}
\star G^1
=
-2\partial_\zeta\partial_{\bar{\zeta}}H\star du,
\ee
or equivalently, $G_{00}=2\partial_\zeta\partial_{\bar{\zeta}}H$. Equation (\ref{eqn-for-profile-funct}) is a well-known equation satisfied by the profile function
on the plane wave-fronts spanned by $\zeta$ and $\bar{\zeta}$ for a given value of the real null coordinate $u$.

\subsection{New massive gravity}

It is well-known that the Einstein tensor leads to no propagating  degrees of freedom at the linearized level for the vacuum field equations in three dimensions.
Recently, Bergshoeff, Hohm and Townsend \cite{BHT,deser-nmg} constructed a particular 3D gravity theory with a massive spin-2 propagating mode, which is often called New Massive Gravity (NMG) theory (and sometimes BHT gravity). The Lagrangian of the NMG  theory composed of  the Einstein-Hilbert term is supplemented by a specific combination of  curvature-squared  terms.  NMG theory is closely related  to  the topologically massive gravity theory \cite{TMG} that follows from the Einstein-Hilbert Lagrangian supplemented by a Lorentz Chern-Simons term in the sense that the topologically massive gravity field equations can be considered as square root of those of the NMG theory \cite{aliev} with the field equations written as a form of a tensorial, Klein-Gordon-type equation.

The four dimensional version of NMG, the so-called ``Critical Gravity" recently introduced  by L\"u and Pope \cite{lu-pope}, involves a Lagrangian with the Einstein-Hilbert term and a cosmological constant extended by quadratic-curvature terms. With the help of fine tuned parameters of the model,  the  massive scalar fields can be eliminated and the massive spin-2 field becomes massless in curved backgrounds. 

Stelle \cite{stelle}  showed that in four spacetime dimensions, the analysis of the linearized field equations for the general quadratic curvature gravity leads to eight degrees of freedom corresponding to massless spin-2, massive spin-2 and a massive scalar mode. The massive modes correspond to the notorious unstable
ghost modes, known as Boulware-Deser  ghost modes \cite{boulware-deser}. In the original derivation presented  by Bergshoeff, Hohm and Townsend, the massive spin-0 mode is eliminated by construction (which amounts to the relation $R_L=0$ in the theory, thanks to the property that the trace of the field equations is of second order) and therefore resulted in  massive spin-2 modes as the only propagating modes about a flat background. 

Recently, de Rham, Gabadadze and Tolley constructed a gravitational theory (dRGT theory) \cite{derham-gabadadze-tolley,derham-gabadadze} of interacting massive spin-2 fields which is not plagued by ghost  modes. Later, the dRGT theory is cast in a convenient form in terms of two sets of coframe 1-form fields
by Hinterbichler  and Rosen \cite{hinterbichler-rosen,hinterbichler-rev,derham} in  a form which is much suited to the spirit of the current work. 
Similarly, the dRGT model formulated in terms of two sets of basis coframe fields and dual connection 1-forms \cite{zwei-dreibein}, unifies the  massive gravity models in three dimensions and eliminates  the Boulware-Deser ghost mode that is a common feature in the nonlinear massive gravity models (See, also \cite{banados}).

The construction of the NMG Lagrangian (\ref{NMG-lag}) starts with a Pauli-Fierz mass term  and a Einstein-Hilbert term together with an auxiliary 
symmetric, second rank tensor field. One can conveniently apply the technique presented in the previous sections to derive the quadratic curvature form of the NMG Lagrangian. The derivation is particularly transparent in the exterior algebra language. The derivation  is valid for any spacetime dimensions $n\geq3$ which, in effect, eliminates the massive scalar mode arising from the quadratic curvature part eventually.

The NMG Lagrangian  involves the gravitational field variables corresponding to the 1-form field $h_a=h_{ab}e^b$ and an auxiliary 1-form field $f_a=f_{ab}e^b$ enjoying the same properties as the perturbation 1-forms $h_a$. The Lagrangian  explicitly reads
\be\label{quadratic-NMG-lag1}
L_{NMG}^{(2)}[h_a,f_a]
=
-
\frac{1}{2}h_a\wdg \star G_L^{a}[h_b]
+
f_a\wdg \star G_L^{a}[h_b]
+
\frac{1}{2}m^2f_a\wdg e_b\wdg\star(f^b\wdg e^a),
\ee
where the first term is the Einstein-Hilbert term, quadratic in the perturbation 1-form $h_a$, whereas the second term on the right hand side can be regarded as a Lagrange multiplier term. $G_L^{a}[f_b]$ stands for the Einstein form linearized in the metric perturbation 1-forms $f_a$ as in (\ref{lin-ein-form2}).
The significance of the specific combination of these two terms in (\ref{quadratic-NMG-lag1}) will be apparent as one proceeds.
The form of quadratic curvature terms given in (\ref{quadratic-NMG-lag1}) is also referred to as ``natural bimetric  form" (see, for example, the recent work 
\cite{hinterbichler2}) for generic four dimensional quadratic curvature models.

By using the expression (\ref{lin-ein-form2}) for the linearized Einstein form, the second term on the right hand side can be rewritten conveniently as
\be
f_a\wdg \star G_L^{a}[h_b]
=
h_a\wdg \star G_L^{a}[f_b]
+
\frac{1}{2}d\left[h_a\wdg e_b\wdg \star\left(df^b\wdg e^a\right)-f_a\wdg e_b\wdg \star\left(dh^b\wdg e^a\right)\right]
\ee
where the total differential on the right  hand side can be disregarded when  it appears in an action integral.
To put it in other words, the second order differential operator implicitly defined by the linearized Einstein form (\ref{lin-ein-form2}) is Hermitian and expressed in terms of the exterior algebra of differential forms, the Hermiticity property readily  follows from the identities satisfied by  the Hodge dual and the exterior derivative operators without identifying the differential operator in terms of partial derivatives. Consequently, the Lagrangian (\ref{quadratic-NMG-lag1})
can be rewritten in the form
\be\label{quadratic-NMG-lag2}
L_{NMG}^{(2)}[h_a,f_a]
=
h_a\wdg 
\star G_L^{a}[f_c-\frac{1}{2}h_c]
+
\frac{1}{2}m^2f_a\wdg e_b\wdg\star(f^b\wdg e^a),
\ee
up to an omitted exact 3-form.
The form of the Lagrangian (\ref{quadratic-NMG-lag2}) allows one to regard $h_a$ as auxiliary variables as well. Thus, $h_a$ can be eliminated from (\ref{quadratic-NMG-lag2}) simply by using the field equations for $h_a$ which explicitly read
\be\label{lin-ein-nmg-f-h}
\frac{\delta L_{NMG}^{(2)}}{\delta h_a}=\star G_L^{a}[f_c-h_c]=0.
\ee
These are  just Einstein's vacuum equations with the Einstein tensor linearized in the 1-forms $f_c-h_c$, and therefore (\ref{lin-ein-nmg-f-h}) are  satisfied identically for $h_a=f_a$. Eventually, using this result to simplify the Lagrangian (\ref{quadratic-NMG-lag2}), one readily recovers the Pauli-Fierz Lagrangian (\ref{mGR-lag}) and by construction, the Lagrangian (\ref{quadratic-NMG-lag1}) is  equivalent to the Pauli-Fierz Lagrangian expressed  in terms of  an auxiliary field symmetric, second rank tensor field variable and a suitable Lagrange multiplier term.

The elimination of the field variable $h_a$ in the  Lagrangian (\ref{quadratic-NMG-lag2}) leads to the fact that it has a massive spin-2 particle content. One can now use the form of the Lagrangian (\ref{quadratic-NMG-lag2}) to  eliminate  the auxiliary fields $f_a$ in the same way as it is done for $h_a$ above by simply calculating the corresponding field equations. Explicitly, the field equations that follow from $\delta L_{NMG}^{(2)}/\delta h_a=0$ can be written  in the form
\be\label{eqns-motion1}
\star G^a_L+m^2\star (f^a-fe^a)=0.
\ee 
Assuming that  the auxiliary fields $f_a$  stand for their linearized form in the field equations, one can rewrite (\ref{eqns-motion1}) without the linear approximation as
\be\label{NMG-f-eliminate}
G^a=-m^2(f^a-fe^a),
\ee
where $f\equiv f^{a}_{\fant{a}a}$, although no label $L$ has been used for the auxiliary 1-form fields $f_a$ up to Eq. (\ref{NMG-f-eliminate}) following the original notation. By calculating the trace of the field equations, one can find that the trace of the auxiliary field variable $f$ is proportional to the scalar curvature as
\be\label{trace-eqn-NMG}
f
=
-\frac{1}{4m^2}R.
\ee
Using (\ref{trace-eqn-NMG}) in the field equations (\ref{NMG-f-eliminate}), one eventually obtains the auxiliary fields in terms of the metric field as 
\be
f^a
=
-\frac{1}{m^2}L^a,
\ee
where $L^a$'s are  Schouten 1-forms $L_a\equiv L_{ab}\theta^b$ defined in terms of the Ricci tensor and scalar curvature as 
\be\label{schouten-tensor-def}
L_{ab}=R_{ab}-\frac{1}{4}\eta_{ab}R.
\ee
Consequently, the nonlinear version of the Lagrangian (\ref{quadratic-NMG-lag1}), with the auxiliary variable $f_a$ eliminated, explicitly reads
\be\label{NMG-lag-3}
L_{NMG}
=
\frac{1}{2}R*1-\frac{1}{m^2}L_a\wdg *G^a+\frac{1}{2m^2}L^a\wdg \theta^b\wdg*\left(L_b\wdg \theta_a\right)
\ee
With the help of the definition (\ref{schouten-tensor-def}) and the identities (\ref{exterior-id2}), one can show that the NMG Lagrangian (\ref{NMG-lag-3}) can be rewritten in a familiar form as
\be\label{NMG-lag}
L_{NMG}
=
\frac{1}{2}R*1-\frac{1}{2m^2}\left(R_{ab}R^{ab}-\frac{3}{8}R^2\right)*1.
\ee
Note that along the derivation of the NMG Lagrangian 3-form, the auxiliary field stands for a tensor as well as its linearization and
the use of exterior algebra makes the manipulations in the  derivation of the new  massive gravity Lagrangian particularly transparent.
It is well-known that the general combination of quadratic curvature terms complementing the Einstein-Hilbert action leads to massive spin-0 
and massive spin-2 modes upon linearization around a flat background. 

In $n$ spacetime dimensions, the above auxiliary fields procedure yields the quadratic curvature part of the form
\be\label{n-dim-secon-order-trace-lag}
\left(R_{ab}R^{ab}-\frac{n}{4(n-1)}R^2\right)*1
\ee
up to an overall constant. For $n=4$, the Lagrangian 4-form in (\ref{n-dim-secon-order-trace-lag}) is equivalent to the conformally invariant  theory, namely the Weyl-squared gravity \cite{bach}.

In a more general geometrical context, Tekin \cite{bayram-tekin-nmg} recently  used the technique introduced in this section
in his study dedicated to the calculation of masses for the spin-0 and spin-2 modes in general quadratic curvature models around a curved background in four dimensions. By making use of the notion of the  equivalent quadratic curvature Lagrangian  
\cite{gullu-sisman-tekin,gullu-sisman-tekin2}, he also obtains the particle spectrum for the gravitational theories  with the general Lagrangian involving  a given function of the Riemann tensor.

\section{Some further development}

The application of the linearization technique described above to a given gravitational model naturally requires the corresponding field equations to be expressed in terms of the exterior algebra of differential forms. This section is devoted to such an extension  of the method to derive linearized field equations in the desired form simply by making use of the variational derivative of a given Lagrangian with respect to the Levi-Civita connection 1-forms.

Let us consider a generic gravitational Lagrangian volume 4-form  $L=L[\omega^{a}_{\fant{a}b}, \theta^b]$, depending on the variables $\omega^{a}_{\fant{a}b}$ and $\theta^a$. For a metric theory these variables are not independent. Furthermore, the local Lorentz invariance of a Lagrangian forbids the explicit appearance of the connection 1-forms in the Lagrangian form because it is not tensorial and therefore, in a metric theory,  the connection can enter into the Lagrangian, for example, through a curvature expression. Likewise, the metric theory can be obtained by the first order formalism \cite{kopczynski,hehl} where $\omega^{a}_{\fant{a}b}$ and $\theta^a$ are assumed to be  independent variables and then constrain  the independent connection 1-forms to be metric compatible  ($\omega_{ab}+\omega_{ba}=0$ relative to an orthonormal coframe) and torsion-free. This can be achieved, for example,  by introducing the corresponding Lagrange multiplier terms to the original Lagrangian. On the other hand, in a direct manner more suitable to linearization, it is possible to convert the variational derivative with respect to  Levi-Civita connection 1-form into a variational derivative with respect to coframe variational derivatives by making use of the variational derivative of the Cartan's structure equations (\ref{se1}). The  method involves the calculation of variational derivative with respect to connection 1-forms and can be described as follows.

 For this purpose,  note first that the variational derivative of the curvature 2-forms can be expressed in general as 
\be\label{var-id-conn-curv}
\delta\Omega^{a}_{\fant{a}b}=D\delta \omega^{a}_{\fant{a}b}
\ee
where $D$ is assumed to be the covariant exterior derivative with respect to a Levi-Civita connection.
Consequently, the terms that turn out to be linear in the metric perturbation  only arise from the variational derivatives of the terms  with respect to the  
connection 1-forms. Therefore, in order to obtain the linearized field equations, it is sufficient to evaluate the variational derivative with respect to connection 1-forms in a suitable manner.  

 The metric field equations can be obtained, as mentioned above, by using the first order formalism which treats  the connection and the coframe 1-forms independently, and the torsion-free condition (\ref{se1}) satisfied by the connection  has to be implemented into the variational derivatives with respect to  the connection 1-forms. Because this is a set of dynamical equations among the variables assumed to be independent initially, one can introduce a Lagrange multiplier term of the form $\lambda_a\wdg \left(d\theta^a+\omega^{a}_{\fant{a}b}\wdg\theta^b\right)$ to impose the constraint (\ref{se1}) on the independent connection. Eventually, one eliminates the Lagrange multiplier forms $\lambda_a$ from the coframe variational derivatives to obtain metric field equations in favor of the remaining variable. In this manner,  terms that are linear in metric perturbation 1-forms turn out to be those  obtained from the Lagrange multiplier term imposing the vanishing torsion constraint on the independent connection.
An immediate technical consequence of this observation is that the  linearized equation is always  a total exterior derivative of a 2-form as a consequence of the metric compatibility $\delta\omega_{ab}+\delta\omega_{ba}=0$ and that $D^L\equiv d$ with flat background. The terms arising from the Lagrange multiplier can practically be found \cite {baykal_epjp} by first noting that the variational derivative of the constraint 
(\ref{se1}) is of the form
\be
D\delta\theta^a+\delta\omega^{a}_{\fant{a}b}\wdg\theta^b=0,
\ee
where $D$ here stands for the covariant exterior derivative with respect to the Levi-Civita  connection.
These equations relating the variational derivatives $\delta \theta^a$ and $\delta\omega^{a}_{\fant{a}b}$
can be inverted to have
\be\label{var-se1}
\delta \omega^{a}_{\fant{b}b}
=
\frac{1}{2}i^ai_b (D\delta \theta^c\wdg \theta_c)-i^a D\delta\theta_b+i_bD\delta \theta^a
\ee
in the same way as the connection 1-forms can be inverted to express them in terms of the exterior  derivatives of the basis coframe 1-forms. 
The expression (\ref{var-se1}) can be used to convert the variational derivatives with respect to the connection 1-forms into the variation derivatives with respect to coframe 1-forms. Consequently, one can use this result and (\ref{lin-basis-coframe}) to have
\be\label{lin-conn-variation}
(\delta\omega^{a}_{\fant{b}b})^L
=
\frac{1}{2}i^ai_b (d\delta h^c\wdg e_c)-i^ad\delta h_b+i_bd\delta h^a
\ee
where, the coframe variational derivatives are expressed in terms of the variation of the metric perturbation 1-forms.
The convenient expression  (\ref{lin-conn-variation}) then can used to identify the coefficient of metric perturbation 1-form $\delta h_a$ as the linearized equations.

Finally, note that the terms linearized in the metric perturbation survives the variational derivative only if the Lagrangian contains terms that are at most 
quadratic in curvature components because, schematically, for a typical Lagrangian $R^n$, a polynomial  in curvature $R$, one has   $\delta R^n\sim (\delta R) R^{(n-1)}\sim \delta g\nabla \nabla R^{(n-1)}$. To obtain the linearized form of field equations following from a Lagrangian involving higher powers of 
curvature  components, G\"ull\"u et al. \cite{gullu-sisman-tekin,gullu-sisman-tekin2} developed a method to convert a higher curvature theory  to an equivalent quadratic curvature theory. In this regard, the linearization of general quadratic curvature models turns out to be  of considerable importance
since their linearization encompasses all higher curvature theories.

In order to illustrate  the details of the above recipe, let us calculate the linearized form of the field equations that follows from the higher order gravitational Lagrangians $L=R^2*1$ and $L=dR\wdg*dR$. 

\subsection{Fourth order gravity with $L=R^2*1$}

In general, the field equations that follow from the gravitational Lagrangian of the form $f(R)*1$ are of fourth order in metric components 
with $f$ being a sufficiently smooth algebraic function of scalar curvature \cite{buchdahl}. The simplest function $f(R)=R^2$ displays the main features of generic $f(R)$ models.

The total variational derivative of the Lagrangian 4-form $L=R^2*1$ can be written  as
\be\label{R2-var1}
\delta L=2R(\delta R)*1+R^2\delta*1,
\ee
where only the first term is of interest for the linearization. Thus, it is more convenient to rewrite the variational derivative (\ref{R2-var1}) in the form
\be\label{R2-var2}
\delta L=2R\delta (R*1)-R^2\delta*1.
\ee
Furthermore, by using $R*1=\Omega_{ab}\wdg *\theta^{ab}$, (\ref{R2-var2}) can be put  into a convenient form as 
\be
\delta L=2R(\delta \Omega_{ab}\wdg *\theta^{ab}+\Omega_{ab}\wdg \delta *\theta^{ab})-R^2\delta*1.
\ee
Evidently, the linear terms arise only from the variational term $2R \delta \Omega_{ab}\wdg *\theta^{ab}$ on the right hand side, which contains  terms that are of fourth order. With the help of the relation (\ref{var-id-conn-curv}), one finds
\be
\delta L=\delta \omega_{ab}\wdg 2D(R*\theta^{ab})+\delta\theta_c\wdg (2R\Omega_{ab}\wdg *\theta^{abc}-R^2*\theta^c).
\ee
Finally, one can rewrite the first term in terms of the linearized quantities with the help of (\ref{lin-conn-variation}) to obtain 
\begin{eqnarray}
\delta L
&=&
(\delta \omega_{ab})^L\wdg 2d(R*\theta^{ab})^L+\ldots
\nonumber\\
&=&
\left(\frac{1}{2}i^ai_b (d\delta h^c\wdg e_c)-i^ad\delta h_b+i_bd\delta h^a
\right)
\wdg
2dR^L\wdg\star e^{ab}+\ldots
\nonumber\\
&=&
\delta h_a\wdg 4d\star (dR^L\wdg e^{a})+\ldots
\end{eqnarray}
Thus, the result can be conveniently expressed in the desired form,  as in the linearized Einstein 3-forms, as
\be\label{lin-R2-form-lang}
\left(\frac{\delta L}{\delta \theta^a}\right)^L
=
(*E^a)^L=\star E^a_L= 4d\star (dR^L\wdg e^{a})
\ee
in terms of the 1-forms $E_a\equiv E_{ab}e^b$. One then uses the result (\ref{lin-R2-form-lang}) to determine the components of $E^L_{ab}$ which can be expressed in the form 
\be\label{lin-R2-coord-lang}
E_{ab}^L
=
4\left(-\partial_a\partial_b R^L+\eta_{ab}\Box R^L\right).
\ee
Once more,  it is worth emphasizing that as in the case of the linearized Einstein field equations, the equation 
(\ref{lin-R2-form-lang}) yields precisely the linearized field equations (\ref{lin-R2-coord-lang}) obtained by the coordinate methods \cite{pechlaner-sexl} 
for $E_{ab}^L$. For completeness, note that the vacuum field equations $*E^a=0$ can be expressed explicitly by using
\be
*E^a
=
4D*(dR\wdg \theta^a)-4R*R^a+R^2*\theta^a.
\ee

Then the linearized form of the field equations (\ref{lin-R2-form-lang}) can be rearranged as
\be
\star E^a_L=2d\left[e_b\wdg i_b\star(dR^L\wdg e^a)\right],
\ee
where the expression $\star E^a[h_b]$ on the right hand side is proportional to $\star G^L_a[R^Le^b]$, that is, the linearized Einstein form with the perturbation 1-forms $h^a$  replaced by $R^Le^a$. The nested structure of the linearized quadratic curvature gravity is readily available by inspection from the field equations using  (\ref{lin-R2-form-lang}).

Yet another interesting observation on the result (\ref{lin-R2-form-lang}) is that the linearized equations may be expressed  as an exact form in flat background. This is not a particular feature of the pure $R^2$ gravity. Consequently, a matter energy momentum $\star T^a_L$ satisifes $d\star T^a_L=0$  
(which reads $\partial_a T^{a}_{\fant{a}b}=0$ in compoent form) as in general relativity. However, there is an additional second order  
identity satisfied by the linearized equations (\ref{lin-R2-form-lang}) which can be derived  as follows 

For simplicity of the argument, let us assume that the field equations with a matter energy-momentum $*T^a$ can be written in the form $*E^a=4*T^a$. Then, the linearized trace  explicitly reads 
\be\label{lin-trace}
3d\star dR^L=-\star T,
\ee
where $T$ stands for the trace of the matter energy-momentum tensor. The trace (\ref{lin-trace}) can also be written in the form $3\Box R^L=-T$.
Now, by applying $\Box$ to the linearized equations (\ref{lin-R2-form-lang}) and taking into account the fact that $\Box$ commutes with $d$ and $\star$, one finds
\be\label{box-lin-eqn}
d\star\left(d\left(\Box R\right)\wdg e^a\right)=\Box T^a_{\fant{a}b}\star e^b.
\ee
Finally, by combining (\ref{lin-trace})  and (\ref{box-lin-eqn}), one finds \cite{pechlaner-sexl}
\be\label{second-order-id}
\eta_{ab}\Box T-\partial_a \partial_b T =3 \Box T_{ab}.
\ee
Evidently, the second order identity (\ref{second-order-id}), which is to be satisfied by any matter energy-momentum tensor, is  valid only in the linear approximation. This peculiar identity for the \emph{pure} quadratic curvature gravity implies that the equations  (\ref{lin-R2-form-lang}) do not admit a solution  with the energy-momentum 1-forms of the form $T^{0}=-\rho(x)e^{0}$, corresponding to a static mass distribution $\rho(x)$ having  spatial extension.

\subsection{Sixth order gravity with $L=dR\wdg *dR$}

Although the sixth order Lagrangian $dR\wdg *dR$ is in a more suitable form for our purposes, note that it  can also be rewritten in the form $-R\Box R*1$ up to an exact form. In this section  we have $\Box\equiv -\left(d^\dagger d+dd^\dagger\right)$ with  $d^\dagger$ standing for the exterior coderivative 
\cite{benn-tucker}. The coderivative can be defined in terms of the exterior derivative and the Hodge dual as $d*=(-1)^p*d^\dagger$. As for the other operators used above, the operators $\Box$ and $d^\dagger$  act on forms defined in curved space as well as those defined in  flat background spacetime. In particular, it follows from the definitions that, $\Box$  acting on a 0-form $\phi$ in flat spacetime yields $\eta^{ab}\partial_a\partial_b\phi$ defined in the previous sections.

The total variational derivative of the Lagrangian can be written in the form
\be
\delta L=2d\delta R\wdg *dR-\delta\theta^a\wdg \left(i_a dR *dR+dRi_a\wdg*dR \right)
\ee
where the second term on the right hand side arises from commuting the variational derivative with the Hodge dual.
The expression on the right hand side can be rearranged to a suitable form  by using the identity $d*=(-1)^{p}*d^\dagger$
acting on a $p$-form  and that the variational derivative commutes with the exterior derivative.
Because one has to consider only the variational derivative with respect to the connection 1-forms,  it is convenient to rewrite the variational derivative in the form
\be
\delta L=2\delta (R*1)\Box R+\ldots
\ee
Now by making use of  $R*1=\Omega_{ab}\wdg *\theta^{ab}$ and subsequently using the variational identity (\ref{var-id-conn-curv}), one can obtain
\be
\delta L=2\delta \omega_{ab}\wdg D (\Box R *\theta^{ab})+\ldots
\ee
where the omitted terms are the variational derivative terms with respect to the coframe basis 1-forms.
As in the previous example, by using now the relation (\ref{lin-conn-variation}), one finally obtains the corresponding linearized equations in the form
\be\label{lin-sixth-order-eqn}
\left(\frac{\delta L}{\delta \theta^a}\right)^L
=
(*E^a)^L=\star E^a_L=-4d\star \left(d(\Box R^L) \wdg e^{a}\right),
\ee
where $\Box$ now refers to the flat background spacetime.

 Note that in this case $\star E^L_a[h_b]$ is proportional to $\star G^L_{a}[\Box R^Le^b]$ and compared to the expression for the linearization of the quadratic curvature gravity, one can deduce that the only difference in the corresponding expressions shows up  as the argument of the Einstein form.

As in the case of the quadratic curvature gravity, pure sixth order gravity coupled to a matter field also leads to an additional second order differential identity in the linearized approximation for a matter energy-momentum tensor. More precisely, if one assumes that pure sixth order gravitational equations can be written in the form $*E^a=4*T^a$ with  matter energy-momentum 1-forms $T^a$, then the linearized trace reads
\be\label{sixth-order-lin-trace}
3d\star d(\Box R^L)=-\star T
\ee
where $T\equiv T^{a}_{\fant{a}a}$ is the trace of the energy momentum tensor.
Equivalently, this equation can also be written in the form $3 \Box^2 R=-T$. Consequently, by inserting the trace expression given in (\ref{sixth-order-lin-trace}) into the equations (\ref{lin-sixth-order-eqn}), one arrives at
\be
d\star \left(dT \wdg e^{a}\right)
=
3\Box T^{a}_{\fant{a}b}\star e^b
\ee
which yields the identity (\ref{second-order-id}) written in component form.
Thus, the result of Pechlaner and Sexl that is rederived in the previous section is also valid here for pure sixth order gravity. This may be an indication of its validity  for a range of general higher order gravitational models, for example, with the Lagrangian of the form $L=R^2*1+dR\wdg*dR$.

Finally, note that the field equations for the sixth order gravity in terms of the exterior algebra differential forms explicitly read
\be 
*E^a
=
-4D*\left(d(\Box R)\wdg \theta^a\right)+4\Box R*R^a-2(i_adR)*dR+i_a(dR\wdg *dR),
\ee
and the general higher order gravitational Lagrangians  involving $R\Box^k R*1$ ($k$ being a positive integer) were studied previously by Schmidt \cite{schmidt}.  Such  Lagrangians leading to $(2k+2)$th order metric equations are shown to be conformally equivalent to Einstein-multi-scalar gravitating models \cite{schmidt2}. More recently, a sixth order gravity where the Einstein-Hilbert term with a cosmological constant term complemented by $R^2$ and $R\Box R$ terms are studied by Bergshoeff et al. \cite{bergshoeff-tricritical-3d-gr} in three dimensions. For the fine-tuned parameters of the model, they showed that  two massive graviton modes become massless in addition to the already existing masless spin-2 modes after linearizing the field equations around an AdS background.

\section{Nonvanishing torsion}

Up to this section, the discussion of linearization  has been confined to  the Riemannian geometry where the torsion and non-metricity vanish identically,
and the metric tensor is fundamental to the underlying  geometrical structure. For the geometrical setting allowing a metric compatible connection with nonvanishing torsion, namely for the Riemann-Cartan geometry \cite{hehl-heyde-kerlick} the connection $\Lambda^{a}_{\fant{a}b}$ is determined by independent field equations for the connection 1-forms coupling to the spin of matter fields. In general, the connection $\Lambda^{a}_{\fant{a}b}$ can be decomposed into its Riemannian and contortion parts
\be\label{RC-decomp}
\Lambda^{a}_{\fant{a}b}=\Omega^{a}_{\fant{a}b}+K^{a}_{\fant{a}b}
\ee
where the contortion 1-forms $K_{ab}=-K_{ba}=K^{a}_{\fant{a}bc}\theta^c$ are related to the torsion 2-forms $\Theta^a$ by 
\be
\Theta^a=K^{a}_{\fant{a}b}\wdg \theta^b,
\ee
The torsion 2-forms in turn can be expressed in terms of the components of the torsion tensor $T^a_{\fant{a}bc}$ as $\Theta^a\equiv \frac{1}{2}T^a_{\fant{a}bc}\theta^{b}\wdg \theta^c$. The curvature 2-forms $\Omega^{a}_{\fant{a}b}(\Lambda)$ corresponding to the Riemann-Cartan connection can be written as  
\be\label{RC-curv-2form-def}
\Omega^{a}_{\fant{a}b}(\Lambda)
=
d\Lambda^{a}_{\fant{a}b}+\Lambda^{a}_{\fant{a}c}\wedge \Lambda^{c}_{\fant{a}b}.
\ee
Furthermore, with the help of the decomposition given in (\ref{RC-decomp}), it is possible to decompose (\ref{RC-curv-2form-def}) in a similar way as
\be\label{RC-curv-decomp}
\Omega^{a}_{\fant{a}b}(\Lambda)
=
\Omega^{a}_{\fant{a}b}(\omega)+D(\omega) K^{a}_{\fant{a}b}
+
K^{a}_{\fant{a}c}\wdg K^{c}_{\fant{a}b}
\ee
where $\Omega^{a}_{\fant{a}b}(\omega)$ are the curvature 2-forms corresponding to the Levi-Civita connection  whereas $D(\omega)$ is the covariant exterior derivative with respect to it. The expressions (\ref{RC-curv-decomp}) then allow one to consider various possibilities for the weak field approximations.
One can consider weak metric fields, and/or weak torsion fields as well. In general, the nonvanishing torsion introduces a term of the form
\be
\left[D(\omega) K^{a}_{\fant{a}b}\right]^L=d (K^{a}_{\fant{a}b})^L
\ee
to linearized Riemann-Cartan curvature 2-forms to first order in the field variables. Consequently, the corresponding linearized Einstein 3-forms take the form
\be\label{lin-RC-ein}
\left[*G^a(\Lambda)\right]^L
=
\star \left[G^a(\omega)\right]^L-\frac{1}{2}d K_{bc}^L\wdg \star e^{abc}
\ee
Note that in this general case, one has two independent weak field 1-forms, $h_a=h_{ab}e^b$ and $K_{ab}^L=K_{abc}^Le^c$ and that the contortion 1-forms $K_{ab}$ are determined by the independent connection equations \cite{arkuszewski}. It is well-known \cite{hehl-heyde-kerlick} that if the equations for the Riemannian connection 1-forms are algebraic, then the metric field equations, and its linearized form (\ref{lin-RC-ein}) can be expressed in terms of Riemannian quantities by eliminating torsion terms.

If the pure gravity action at hand is linear in the curvature, (e.g., the  Einstein-Hilbert or the Brans-Dicke action \cite{bd-original,dereli-tucker-plb1982}) both the zero-torsion constrained and the unconstrained variations lead to the same set of linearized  vacuum field equations. If the action is quadratic or higher order in the curvature, such an equivalence does not occur in general.

Finally, it is interesting to note that the linearized form of the field equations for the teleparallel gravity \cite{mielke1,mielke2,saridakis}, where one has vanishing curvature but a  nonvanishing torsion, has been considered only relatively recently  by Obukhov and Pereira \cite{obukhov-pereira}.

\section{Concluding comments}

An interesting remark concerning  the linearization presented above is the following. By inserting the connection expression (\ref{inverted-se1}) into the reduced Einstein-Hilbert Lagrangian 4-form (\ref{EH-lag2}), the Einstein-Hilbert Lagrangian can be expressed solely in terms of basis 1-forms in the form 
\be\label{EH-coframe-form}
L_{EH}
=
-
\frac{1}{2}d\theta^a\wdg \theta^b\wdg *\left(d\theta_b\wdg \theta_a\right)
+
\frac{1}{4}d\theta^a\wdg \theta_a\wdg *\left(d\theta^b\wdg \theta_b\right)
-
d(\theta_a\wdg *d\theta^a)
\ee
by eliminating the connection 1-forms in favor of the basis coframe 1-forms \cite{thirring}. 
The Pauli-Fierz Lagrangian (\ref{red-EH-lag}) is obtained by inserting (\ref{lin-basis-coframe}) into the expression (\ref{EH-coframe-form}) 
where the second term on the right hand side drops out as a consequence of the assumption that $h_{ab}=h_{ba}\Leftrightarrow h_a\wdg e^a=0$.
On the other hand, the remaining  term turns out to be  the germane part of the Einstein-Hilbert Lagrangian in the sense that the full $L_{EH}$ can be recovered 
from it \cite{deser,straumann-talk} by an infinite number of iterations starting from the linearized approximation.

There are several directions that  the linearization technique above can be extended. The above formalism can be applied to the linearization of any modified gravitational field equations around a flat background, which will presumably simplify  more complicated field equations \cite{deser-tekin3,baykal} as well. Such a scheme additionally requires that the field equations are expressed in the language of the exterior algebra of differential forms
as in the case of the quadratic curvature and the sixth order gravitational models studied above.  The calculations for the particular higher order gravitational models studied above imply that the linearization of the gravitational field equations that follows from a Lagrangian  involving only the powers (and the derivatives) of the curvature tensor leads to the mathematical structure similar to the linearized Einstein tensor which may be considered as a general feature that is an imprint of the  curvature tensor.

In this paper we have considered only the flat background and in a physically important direction of development, one can tackle the technical details of  the linearization around an arbitrary curved background \cite{abbott-deser,deser-tekin3,deser-tekin1,deser4,deser-tekin2,cebeci-sarioglu-tekin}. The extension of the linearization technique above is essential in the linearization of  modified gravitational models involving, for example, the general quadratic curvature gravity which admits maximally symmetric vacuum solutions. In this case one has to take  both the zeroth and first order terms in the perturbation 1-forms $h_{a}$ into  account in the linearization calculations as, for example, displayed in the expansions (\ref{basis_expansion}) and (\ref{hodge_basis_expansion}).

\end{document}